\documentclass[review]{elsarticle}
\usepackage{amsmath,amssymb,amsbsy}
\usepackage{lineno}
\usepackage{hyperref}
\usepackage[numbers]{natbib}
\usepackage[T1]{fontenc} 
\usepackage{subcaption}
\modulolinenumbers[5]
\usepackage{tabularx}
\usepackage{wrapfig}
\usepackage{gensymb}
\usepackage{pifont}
\usepackage{tabu}
\usepackage{mathtools}
\usepackage{xcolor}
\usepackage{tikz}
\usepackage{array}
\usepackage{adjustbox}
\usepackage{wasysym}

\newcommand{\form}{$\Delta E_{\mathrm{vac}}^f$}
\newcommand{\ads}{$\Delta E_{\mathrm{ads}}^f$}
\newcommand{\bind}{$\Delta E_{\mathrm{b}}$}
\newcommand{\etdsub}{$E_{\mathrm{2D+S}}$}
\newcommand{\etdtd}{$E_{\mathrm{3D}}$}
\newcommand{\ntdtd}{$N_{\mathrm{3D}}$}
\newcommand{\etd}{$E_{\mathrm{2D}}$}
\newcommand{\ntd}{$N_{\mathrm{2D}}$}
\newcommand{\esub}{$E_{\mathrm{S}}$}
\newcommand{\tikzcircle}[2][red,fill=red]{\tikz[baseline=-0.5ex]\draw[#1,radius=#2] (0,0) circle ;}

\begin{document}

\begin{frontmatter}

    \title{Computational Synthesis of 2D Materials: A High-throughput Approach to Materials Design} 
    \author[1]{Tara M. Boland}
    \author[2]{Arunima K. Singh\corref{cor1}}
    \address[1]{School for Engineering of Matter, Transport and Energy, Arizona State University, Tempe, AZ, USA}
    \address[2]{Department of Physics, Arizona State University, Tempe, AZ, USA}
    \cortext[cor1]{Corresponding author}
    \ead{arunimasingh@asu.edu}
    
    \begin{abstract}
        2D materials find promising applications in next-generation devices, however, large-scale, low-defect, and reproducible synthesis of 2D materials remains a challenging task. To assist in the selection of suitable substrates for the synthesis of as-yet hypothetical 2D materials, we have developed an open-source high-throughput workflow package, $Hetero2d$, that searches for low-lattice mismatched substrate surfaces for any 2D material and determines the stability of these 2D-substrate heterostructures using density functional theory (DFT) simulations. $Hetero2d$ automates the generation of 2D-substrate heterostructures, the creation of DFT input files, the submission and monitoring of computational jobs on supercomputing facilities, and the storage of relevant parameters alongside the post-processed results in a MongoDB database. We demonstrate the capability of $Hetero2d$ in identifying stable 2D-substrate heterostructures for four 2D materials, namely $2H$-MoS$_2$, $1T$- and $2H$-NbO$_2$, and hexagonal-ZnTe, considering 50 cubic elemental substrates. We find Cu, Hf, Mn, Nd, Ni, Pd, Re, Rh, Sc, Ta, Ti, V, W, Y, and Zr substrates sufficiently stabilize the formation energies of these 2D materials, with binding energies in the range of $\sim$0.1 -- 0.6 eV/atom. Upon examining the $z$-separation, the charge transfer, and the electronic density of states at the 2D-substrate interface, we find a covalent type bonding at the interface which suggests that these substrates can be used as contact materials for the 2D materials. \href{https://github.com/cmdlab/Hetero2d}{Hetero2d} is available on GitHub as an open-source package under the GNU license.
    \end{abstract}
    \fntext[fn1]{footnote}
    
\end{frontmatter}

\section{Introduction}
    The emergence of atomically thin, single-layer graphene spawned a new class of materials, known as two-dimensional (2D) materials~\cite{Xu2013, Novoselov2011}. These extraordinary 2D materials have attracted significant attention within the scientific community due to their wide range of properties - from large band-gap insulators to the very best conductors, the mechanically tough to soft and malleable, and semi-metals to topologically insulating~\cite{Singh2015, Paul2017,Blonsky2015,Akiyama2021}. The diverse pool of properties that 2D materials possess promise many novel next-generation device applications in nanoelectronics, quantum computing, field-effect transistors, microwave and terahertz photonics, and catalysis~\cite{Rode2017, Xu2015, Yu2014, Kang2013, Amani2014, Li2019, Luo2016, Yu2014}. Despite the excitement surrounding these promising materials, surprisingly few 2D materials are used in the industry. Roughly 55 of the $>$5,000 theoretically predicted 2D materials have been experimentally synthesized~\cite{Mounet2018, Ashton2017, c2db, Singh2014, Zhou2019}.
   
    Of the various methods used to synthesize 2D materials, substrate-assisted methods such as chemical vapor deposition result in large-area and low-defect flakes at a reasonable cost per mass~\cite{Novoselov2012}. Substrate-assisted methods have the added benefit of being able to synthesize 2D materials that have non-van der Waals (vdW) bonded bulk counterparts. On the other hand, exfoliation techniques, like mechanical exfoliation~\cite{Singh2015}, can only be used to generate 2D flakes from vdW-bonded bulk counterparts. Currently, substrate-assisted synthesis of 2D materials rely on expensive trial-and-error processes requiring significant experimental effort and intuition for choosing the substrate, precursors, and the growth conditions (substrate temperatures, growth rate, etc.) to synthesize 2D materials resulting in the slow progress to realize and utilize these materials. Furthermore, the properties of 2D materials can be dramatically altered by placing them on substrates. For example, the mobility of carriers in 2D-MoS$_2$ is reduced by more than an order of magnitude by placing it on a sapphire substrate~\cite{singh2015al2o3}. To enable the functionalization and to assist in the selection of substrates for synthesis, a detailed understanding of the substrate-assisted modification of energetic, physical, and electronic properties of 2D materials is required. 
    
    In this work, we present the $Hetero2d$ workflow package inspired by existing community workflow packages. $Hetero2d$ is tailored to address scientific questions regarding the stability and properties of 2D-substrate heterostructured materials. $Hetero2d$ provides automated routines for the generation of low-lattice mismatched heterostructures for arbitrary 2D materials and substrate surfaces, the creation of vdW-corrected density-functional theory (DFT) input files, the submission and monitoring of simulations on computing resources, and the post-processing of the key parameters to compute, namely, (a) the interface interaction energy of 2D-substrate heterostructures, (b) the identification of substrate-induced changes in the interfacial structure, and (c) charge doping of the 2D material. The 2D-substrate information generated by our routines is stored in a MongoDB database tailored for 2D-substrate heterostructures.
    
    As an example, we demonstrate the use of $Hetero2d$ in screening for substrate surfaces that stabilize the following four 2D materials - $2H$-MoS$_2$, $1T$- and $2H$-NbO$_2$, and hexagonal-ZnTe. We considered the low-index planes of a total of 50 cubic metallic materials as potential substrates. Using the $Hetero2d$ workflow, we determine that Cu, Hf, Mn, Nd, Ni, Pd, Re, Rh, Sc, Ta, Ti, V, W, Y, and Zr substrates sufficiently stabilize the formation energies of these 2D materials, with binding energies in the range of $\sim$0.1 -- 0.6 eV/atom. Upon examining the $z$-separation, the charge transfer, and the electronic density of states at the 2D-substrate interface using post-processing tools of $Hetero2d$, we find a covalent type bonding at the interface, which suggests that these substrates can be used as contact materials. \href{https://github.com/cmdlab/Hetero2d}{Hetero2d} is shared on GitHub as an open-source package under the GNU license. 
    
\section{DFT Approach to Identifying Stable 2D-Substrate Heterostructures}
    2D materials are inherently meta-stable materials and are often created by peeling 2D films from layered, vdW bonded bulk counterparts. Their meta-stability arises from the removal of the vdW bonds between the individual flakes. However, the vdW bonds are an order of magnitude weaker than the in-plane covalent or ionic bonds of 2D materials, thus many 2D materials can remain stable at room temperature or above. A quantitative measure of the stability of 2D materials to remain as a free-standing 2D film is given by the formation energy, \form, with respect to the bulk phase
    \begin{equation}
    	\label{eq:Eform}
      	\begin{aligned}[t]
    		\hspace*{-1.5cm} \Delta E_{\mathrm{vac}}^f &= \dfrac{ E_{\mathrm{2D}}}{ N_{\mathrm{2D}} } - \dfrac{E_{\mathrm{3D}}}{N_{\mathrm{3D}}},\\
    	\end{aligned}
    \end{equation} where \etd\ is the energy of a 2D material in vacuum, \etdtd\ is the energy of the bulk counterpart of the 2D material, and \ntd\ and \ntdtd\ are the number of atoms in the unit cell of 2D and bulk counterpart, respectively. 
    
    The \form\ of a 2D material indicates the stability of a 2D flake to retain the 2D form over its bulk counterpart, where the higher the \form, the larger the driving force to lower the free energy. Singh et. al. and others have shown that when the \form\ < 0.2 eV/atom, the 2D materials are stable as a free-standing film, but for larger \form's they are highly unstable and may only be synthesized using substrate-assisted methods~\cite{Singh2015, c2db}. 
    
    For substrate surfaces to stabilize a 2D material during the growth processes, the 2D-substrate heterostructure should be energetically stable. Thus the interactions between the 2D material and substrate surface have to be attractive in nature. This interaction energy known as the binding energy can be estimated as, $\Delta E_{\mathrm{b}} = (E_{\mathrm{2D}} + E_{\mathrm{S}} - E_{\mathrm{2D+S}} )/N_{\mathrm{2D}}$, where \etdsub\ is the energy of the 2D material adsorbed on the surface of a substrate, \esub\ is the energy of the substrate slab, \etd\ is the energy of the free-standing 2D material, and \ntd\ is the number of atoms in the unit cell of the 2D material. Note, strain is applied to the 2D material to place it on the substrate surface due to the lattice-mismatch between the two lattices. For the 2D-substrate heterostructure interaction to be attractive, the \bind\ > 0. In addition, this \bind\ should be greater than the \form\ of 2D materials to ensure that the 2D materials remain in their 2D form on the substrate. Singh et. al. has shown previously that the successful synthesis of a 2D material on a particular substrate surface is feasible when the adsorption formation energy, \ads\ = \form\ - \bind\ < 0.

\section{Hetero2d: The High-Throughput Implementation of the DFT Approach}
    \subsection{Introduction}
        The $Hetero2d$ package is an all-in-one workflow approach to model the heterostructures formed by the arbitrary combinations of 2D materials and substrate surfaces. $Hetero2d$ can calculate the \form, \bind, and \ads\ for each 2D-substrate heterostructure and store the relevant simulation parameters and post-processing in a queryable MongoDB database that can be interfaced to and accessed by an application programming interface (API) or a web-portal. $Hetero2d$ is written in Python 3.6, a high-level coding language widely used on modern scientific computing resources. $Hetero2d$ utilizes \textit{MPInterfaces}~\cite{Mathew2016} routines and the robust high-throughput computational tools developed by the Materials Project~\cite{atomate,Jain2013,Jain2015,Ong2013} (MP), namely \textit{atomate}, \textit{FireWorks}, \textit{pymatgen}, and \textit{custodian}.
    
\begin{figure}[!th]
            \centering
            \includegraphics[width=\textwidth]{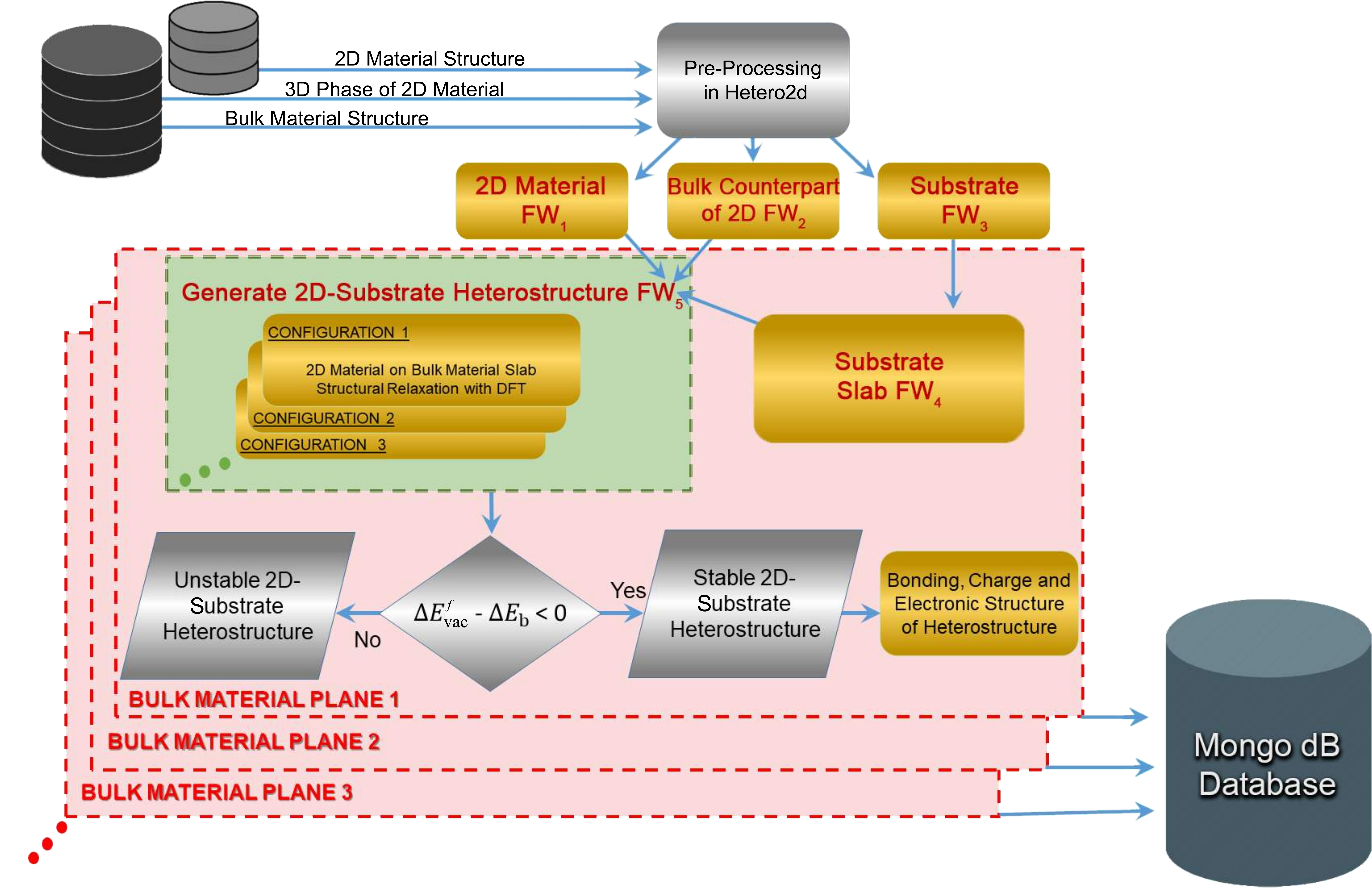}
            \caption{Outline for our computational workflow used in our study to investigate the properties of the 2D-substrate heterostructures as coded in the $Hetero2d$ package. All structures imported from an external database are relaxed using vdW-corrected DFT with our parameters (discussed below) to maintain consistency. Boxes in gold denote a DFT simulation step and boxes in silver denote a pre-processing or post-processing step.}
            \vspace{-0.25\intextsep} 
            \label{fig:Figure1}
        \end{figure}

        $Hetero2d$'s framework is inspired by \textit{atomate}'s straightforward statement-based workflow design to perform complex materials science computations with pre-built workflows that automate various types of DFT calculations. Figure \ref{fig:Figure1} illustrates the framework of our workflow within the $Hetero2d$ package. $Hetero2d$ extends some powerful high-throughput techniques available in existing community packages and combines them with new routines created for this work to generate 2D-substrate heterostructures, perform vdW-corrected DFT calculations, store the stability related data within a queryable database, and analyze key properties of the heterostructure. In the following sections, we discuss each step outlined in Figure \ref{fig:Figure1} underscoring the new computational tools developed for $Hetero2d$.
        
    \subsection{Workflow Framework}
        $Hetero2d$'s \textit{atomate}-inspired framework utilizes the \textit{FireWorks} package to break down and organize each task within a workflow. Workflows within the \textit{FireWorks} package are organized into three task levels -- (1) workflow, (2) firework, and (3) firetask. A workflow is a set of fireworks with dependencies and information shared between them through the use of a unique specification file that determines the order of execution of each firework (FW) and firetask. Each FW is composed of one or more related firetasks designed to accomplish a specific task such as DFT structure relaxation. Firetasks are the lowest level task in the workflow. Firetasks can be simple tasks such as writing files, copying files from a previous directory, or more complex tasks such as calling script-based functions to generate 2D-substrate heterostructures, starting and monitoring a DFT calculation, or post-processing a DFT calculation and updating the database. 
        
        $Hetero2d$'s workflow \textit{get\_heterostructures\_stabilityWF} shown in Figure \ref{fig:Figure1}, has a total of five firework steps (1) FW$_1$: the DFT structural optimization of the 2D material, (2) FW$_2$: the DFT structural optimization of the bulk counterpart of the 2D material, (3) FW$_3$: the DFT structural optimization of the substrate, (4) FW$_4$: the creation and DFT structural optimization of the substrate slab, and (5) FW$_5$: the generation and DFT structural optimization of the 2D-substrate heterostructure configurations. Each firework can be composed of a single or many related firetasks. The tasks are gathered from the specification file that controls the execution of each firetask. For example, FW$_1$ is used to perform a vdW-corrected DFT structure optimization of the 2D material. Note that the DFT simulations are performed using the Vienna \textit{ab initio} simulation package ~\cite{Kresse5, Kresse4, Kresse1, Kresse2, Kresse3}. FW$_1$ is composed of firetasks which (1) write VASP input files to the job's launch directory, (2) write the structure file, (3) run VASP using \textit{custodian}~\cite{Ong2013} to perform just-in-time job management, error checking, and error recovery, (4) collect information regarding the location of the calculation and update the specification file, and (5) perform analysis and convergence checks for the calculation and store all pre-defined information about the calculation in our MongoDB database. A more detailed explanation of each firework in the workflow is discussed in section 3.6, \textit{Workflow Steps}. 
        
    \subsection{Package Functionalities}
        As mentioned earlier, $Hetero2d$ adapts and extends existing community packages to assess the stability of 2D-substrate heterostructures. Table \ref{tab:Table1} lists the functionalities of $Hetero2d$ compared with two other workflow-based packages, \textit{MPInterfaces}~\cite{Mathew2016} and \textit{atomate}~\cite{atomate}, highlighting new and common features within the three packages. 
        \begin{table}
            \centering
            \caption{A list of functionalities present in the $Hetero2d$ package compared with two other workflow-based packages \textit{MPInterfaces} and \textit{atomate}. $Hetero2d$ is the only workflow package with all the specific features needed to create 2D-substrate heterostructures using high-throughput computational methods.}
            \begin{adjustbox}{width=0.5\textwidth}
                \begin{tabular}{|c|c|c|c|}
                \hline
                     & $Hetero2d$ & \textit{MPInterfaces} & \textit{Atomate} \\
                \hline
                    Structure processing & \checked & \checked & \checked \\
                \hline
                    Error recovery       & \checked & \checked & \checked \\
                \hline
                    Database integration & \checked & \checked & \checked \\
                \hline
                    \textit{FireWorks} compatible & \checked &   & \checked \\
                \hline
                    2D hetero. routines  & \checked & \checked &   \\
                \hline
                    2D hetero. workflow  & \checked &   &   \\
                \hline
                    2D post-processing   & \checked &   &   \\
                \hline
                \end{tabular}
            \end{adjustbox}
            \label{tab:Table1}
        \end{table}
        
        All three packages utilize the \textit{pymatgen} package to perform various structure processing tasks. \textit{Pymatgen} is used to perform various types of structure-manipulation processes such as reducing/increasing simulation cell size, creating a vacuum, or creating a slab during the execution of the workflow. Throughout $Hetero2d$, we utilized \textit{pymatgen} to handle structure-manipulation for (a) the bulk materials and (b) some basic pre-/post-processing of structures and generation of files for the DFT calculations. Within $Hetero2d$, \textit{pymatgen}'s structure-manipulation tools are used to create conventional unit cells for the substrate and create the substrate slab surface. Additionally, we have integrated \textit{pymatgen}'s structure analysis modules to decorate the fireworks in the workflow with structural information for each input structure to populate our database. The pre-processing enables one to differentiate crystal phases with similar compound formulas, easily reference and sort data within the database, and perform analysis in later fireworks. 
        
        All three packages use the \textit{custodian} package~\cite{Ong2013} to perform error recovery. Error recovery routines are pivotal for any workflow package to reduce the need for human intervention and correct simple run-time errors with pre-defined functions. Additionally, \textit{custodian} alerts the user if an unrecoverable error has occurred.
        
        Database integration is another functionality present in all three packages that stores and analyzes the vast amount of information generated by each calculation. 
       
        Only $Hetero2d$ and \textit{atomate} are \textit{FireWorks} compatible whereas, \textit{MPInterfaces} uses the python package \textit{fabric} to remote launch jobs over SSH. \textit{FireWorks} is a single package used to define, manage, and execute scientific workflows with built-in failure-detection routines capable of concurrent job execution and remote job tracking over an arbitrary number of computing resources accessible from a clean and flexible Python API. 
       
        Routines used to automate the generation of 2D-substrate heterostructures given user constraints are available in $Hetero2d$ and \textit{MPInterfaces}. \textit{MPInterfaces} implements a mathematical algorithm developed by Zur et. al.~\cite{Zur1984} for generating supercells of lattice-matched heterostructures given two arbitrary lattices and user-specified tolerances for the lattice-mismatch and heterostructure surface area. $Hetero2d$ incorporates functions from \textit{MPInterfaces} to create 2D-substrate heterostructures and enable our package to utilize \textit{FireWorks} which \textit{MPInterfaces} is currently incompatible with. Additionally, by incorporating these routines in $Hetero2d$, we can modify the function to return critical information regarding the 2D-substrate heterostructures that are not returned by the \textit{MPInterfaces} function. Our 2D-substrate heterostructure function returns the strain of the 2D material along \textbf{a} and \textbf{b} lattice vectors, angle mismatch between the \textbf{ab} lattice vectors of the substrate and the 2D material, and scaling matrix used to generate the aligned the 2D-substrate heterostructures. 
        
       
        The 2D-substrate heterostructure workflow and post-processing routines are uniquely available in $Hetero2d$. The workflow automates all steps needed to study 2D-substrate heterostructure stability and properties via the DFT method. The post-processing routines enable a curated database to view all calculation results and perform additional analysis or calculations. 
    
    \subsection{Default Computational Parameters}
        \textit{CMDLInterfaceSet} is based on \textit{pymatgen}'s \textit{VASPInputSet} class that creates custom input files for DFT calculations. Our new class \textit{CMDLInterfaceSet} has all the functionality of the parent \textit{pymatgen} class but is tailored to perform structural optimizations of 2D-substrate heterostructures and implements vdW-corrections, on-the-fly dipole corrections for slabs, generation of custom $k$-point mesh grid density, and addition of selective dynamics tags for the 2D-substrate structures. All DFT calculations are performed using the projector-augmented wave method as implemented in the plane-wave code VASP~\cite{Kresse5, Kresse4, Kresse1, Kresse2, Kresse3}. The vdW interactions between the 2D material and substrate are modeled using the vdW–DF~\cite{Rydber2003} functional with the optB88 exchange functional~\cite{Klimes2011}. 
    
        The \textit{CMDLInterfaceSet} has a default energy cutoff of 520 eV used for all calculations to ensure consistency between structures that have the cell shape and volume relaxed and those that only have ionic positions relaxed. The default $k$-point grid density was automated using \textit{pymatgen}~\cite{Ong2013} routines to 20 $k$-points/unit length by taking the nearest integer value after multiplying $\frac{1}{\textbf{a}}$ and $\frac{1}{\textbf{b}}$ by 20. These settings were sufficient to converge all calculations to a total force per atom of less than 0.02 eV/\AA. Additional information regarding default settings set in the \textit{CMDLInterfaceSet} and convergence tests performed to benchmark our calculations are in the section 1 and 2 of the SI.
    
    \subsection{Workflow Initialization and Customization}
        To use $Hetero2d$'s workflow, \textit{get\_heterostructures\_stabilityWF}, we import the 2D structure, its bulk counterpart, and the substrate structure from existing databases through their APIs. When initialized, the workflow can accept up to three structures (1) the 2D structure, (2) the bulk counterpart of the 2D structure, and (3) the substrate structure in the bulk or slab form. 
        
\begin{wrapfigure}[11]{r}{0.55\textwidth}
            \vspace{-1.4\intextsep} 
            \hspace*{-0.4\columnsep}\includegraphics[width=0.55\textwidth]{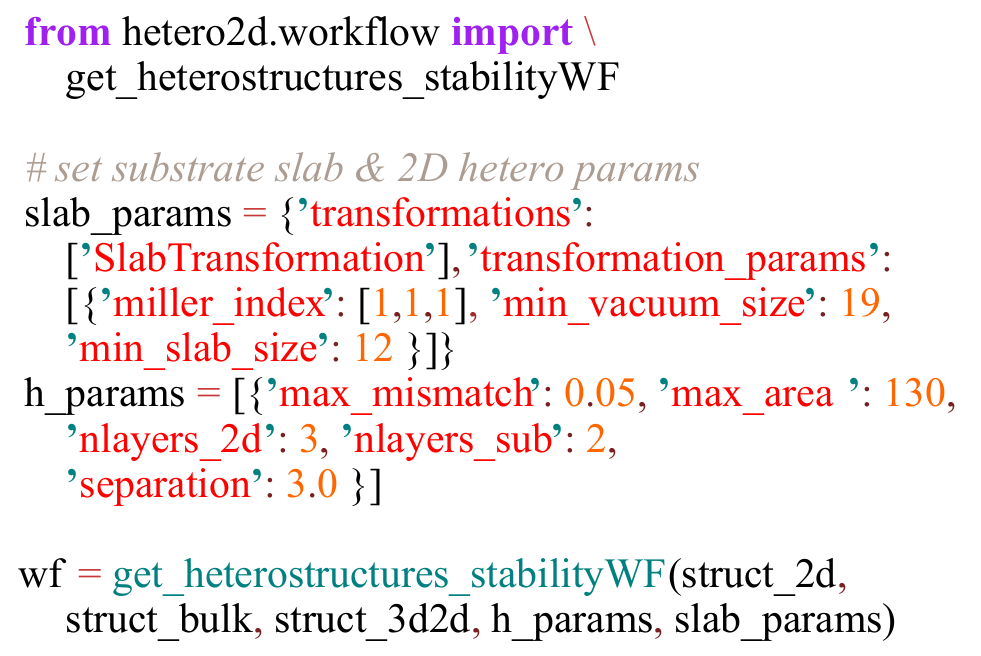}
            \vspace{-0.55\intextsep} 
            \caption{Simplified workflow illustrating the setup necessary to setup the 2D-substrate heterostructure workflows using \textit{get\_heterostructures\_stabilityWF} used throughout this work. A full example jupyter notebook is located in the SI.}
            \vspace{10\intextsep}
            \label{fig:Figure2}
        \end{wrapfigure}

        To perform structure transformations to generate the substrate slabs or the 2D-substrate heterostructures, our workflow requires two dictionaries during initialization -- the (1) \textit{h\_params} and (2) \textit{slab\_params} dictionary. Figure \ref{fig:Figure2} is a code excerpt demonstrating the parameters one can supply to generate a 2D-substrate heterostructure on a (111) substrate slab surface. In Figure \ref{fig:Figure2}, \textit{slab\_params} dictionary generates a substrate slab with a vacuum spacing of 19 \AA\ and a substrate slab thickness of at least 12 \AA. The \textit{h\_params} dictionary creates the lattice-matched, symmetry-matched 2D-substrate heterostructures with 3.0 \AA\ $z$-separation distance between the 2D material and the substrate surface. The \textit{h\_params} dictionary also sets the maximum allowed lattice-mismatch along \textbf{ab} to be less than 5\%, a surface area less than 130 \AA$^2$, sets the selective dynamics tags in the DFT input file to relax all layers of the 2D material and top two layers of the substrate slab. 
        
        The workflow has commands for two VASP executables compiled that incorporate vdW-corrections for performing DFT calculations for (1) 2D materials and (2) 3D materials. The first executable is a custom executable to relax 2D materials with a large vacuum and prevent the vacuum from shrinking by not letting the cell length change in the direction of vacuum spacing. The second executable allows the cell volume to change in all directions. Other optional arguments used to initialize the workflow include dipole correction for substrate slabs, tags for database entries, and avenues to modify the INCAR of each firework in the workflow. The parameters $vis$ and \textit{vis\_i} where $i$=2d, 3d2d, bulk, trans, and iface are used to override the default \textit{VaspInputSet} with one provided by the user. This can be provided for all fireworks using \textit{vis} or for a specific firework using \textit{vis\_i}. The parameters \textit{uis} and \textit{uis\_i} can be set to change the default settings in the INCAR. The parameter \textit{uis} will set the specified parameters for all INCARs in the workflow, while \textit{uis\_i} will set the INCAR parameters for the corresponding firework. Additional details regarding workflow customization options and current functionality available in \textit{Hetero2d} are discussed in SI section 3 as well as an example jupyter notebook.
    
    \subsection{Workflow Steps}
        As mentioned previously, our workflow has five firework steps. Here, we discuss the pre-processing steps that occur when initializing the workflow, each firework, and the firetasks composing each firework for the 2D-substrate heterostructure workflow introduced in section 3.2, \textit{Workflow Framework}.
        
        The first firework, FW$_1$, in the workflow optimizes the 2D material structure. During initialization of the workflow, the 2D material is centered within the simulation cell, obtaining crystallographic information regarding the structure, the \textit{CMDLInterfaceSet} is initialized to create VASP input files, and a list of user-defined/default tags are created for the 2D material. The structure, tags, and \textit{CMDLInterfaceSet} are used to initialize the firework \textit{HeteroOptimizeFW} that performs the structure optimization. The default tags appended to the firework are the unique identification tags (provided to the workflow by the user), the crystallographic information, workflow and firework name, and the structure's composition. In FW$_1$, \textit{HeteroOptimizeFW} executes firetasks that -- (a) create directories for the firework, (b) write all input files initialized using \textit{CMDLInterfaceSet}, (c) submit the VASP calculation to supercomputing resources to perform full structure optimization and monitor the calculation to correct errors, (d) run our \textit{HeteroAnalysisToDb} class to store all information necessary for data analysis within the database, and (e) lastly pass the information to the next firework. Details regarding \textit{HeteroAnalysisToDb} can be found in the next section.
        
        Similar to FW$_1$, FW$_2$ and FW$_3$ perform a full structural optimization for the bulk counterpart of the 2D material and the substrate, respectively. FW$_2$ and FW$_3$ differ from FW$_1$ only in the pre-processing steps. The step to center the 2D material is not performed, however, the conventional standard structure is utilized during the pre-processing for FW$_3$.
    
        FW$_3$ spawns a child firework passing the optimized substrate structure to FW$_4$ which transforms the conventional unit cell of the substrate into a substrate slab using the \textit{slab\_params} dictionary and performs the structure optimization. When the workflow is initialized, FW$_4$ undergoes similar pre-processing steps that are used to initialize the firework \textit{SubstrateSlabFW} that creates a substrate slab from the substrate. \textit{SubstrateSlabFW} is the firework that transforms the conventional unit cell of the substrate into a slab, sets the selective dynamics tags on the surface layers, and sets the number of compute nodes necessary to relax the substrate slab. The \textit{slab\_params} variable is the input dictionary that initializes \textit{pymatgen}'s \textit{SlabTransformation} module that creates the substrate slab. All required and optional input arguments used in the \textit{SlabTransformation} module must be supplied using this dictionary (key: value) format. This dictionary format is implemented to enable $Hetero2d$ to be flexible and extendable in future updates. Additionally, the \textit{slab\_params} dictionary is only required when creating a new substrate slab from a substrate. 
    
        After the first four fireworks have been completed and successfully stored in the database, the fifth firework (FW$_5$) obtains the optimized structures and information from previous fireworks and the specification file. FW$_5$ calls the \textit{GenHeteroStructuresFW} firework to generate the 2D-substrate heterostructure configurations using \textit{h\_params} and spawns a firework to perform structure optimization for each configuration. The input required for the \textit{h\_params} dictionary are those that are required by $Hetero2d$'s \textit{hetero\_interfaces} function. This function attempts to find a matching lattice between the substrate surface and the 2D material. The parameters used to initialize \textit{hetero\_interfaces} are listed in the \textit{h\_params} dictionary shown in Figure \ref{fig:Figure2} and the jupyter notebook in the SI. 
        
        Our function \textit{hetero\_interfaces} generates the 2D-substrate heterostructure configurations utilizing \textit{MPInterfaces}'s interface matching algorithm. We developed \textit{hetero\_interfaces} to ensure functions within the workflow are compatible with \textit{FireWorks}. Additionally, we can return key variables regarding the interfacing matching algorithm, such as the strain or angle mismatch, and store these values in our database. \textit{MPInterfaces} is used to (a) generate heterostructures within an allowed lattice-mismatch and surface area of the supercell at any rotation between the 2D material and bulk material surface and (b) create distinct configurations in which the 2D material can be placed on the bulk material surface based on the Wyckoff positions of the near-interface atoms.
        
        FW$_5$ calls \textit{GenHeteroStructuresFW} which generates the 2D-substrate heterostructure configurations, the total number of configurations is computed, each unique configuration is labeled from 0 to $n$-1, where $n$ is the total number of configurations, and stored under the \textit{Interface Config} tag. For each configuration, a new firework is spawned to optimize each 2D-substrate heterostructure configuration. The data generated within FW$_5$ is stored in the database.
    
        After all previous FWs have successfully converged, \textit{HeteroAnalysisToDb} is called one final time to compute the \form, \bind, and \ads\ for each heterostructure configuration generated by the workflow. The calculation of the \form\ references the simulation for the 2D material and its bulk counterpart. The bulk counterpart is simulated using a standard periodic simulation cell. The calculation of \bind\ references the 2D material, substrate slab, and 2D-substrate heterostructure simulations which all employ a standard supercell slab model. The calculation of the \ads\ references both \bind\ and  \form. Once each value is computed, all the information is curated and stored in the MongoDB database.

    \subsection{Post-Processing Throughout Our Workflow} 
        After each VASP simulation is complete, post-processing is performed within the calculation directory using our \textit{HeteroAnalysisToDb} class, an adaptation of \textit{atomate}'s \textit{VaspToDb} module. It is used to parse the calculation directory, perform error checks, and curate a wide range of input parameters and quantities from calculation parameters and output, energetic parameters, and structural information for storage in our MongoDB. \textit{HeteroAnalysisToDb} detects the type of calculation performed within the workflow and parses the calculation accordingly. \textit{HeteroAnalysisToDb} has the same functionally as \textit{VaspToDb} with additional analyzers developed for 2D-substrate heterostructures that -- (a) identify layer-by-layer interface atom IDs for the substrate and 2D material, (b) store the initial and final configuration of all structures, (c) compute the \form, \bind, and \ads, (d) store the results obtained from the interface matching, and (e) ensure each database entry has any custom tags added to the database such as those appended by the user. The workflow design ensures that the DFT simulations for each 2D-substrate surface pair will be performed independently of each other, but as soon as all simulations are completed for each 2D-substrate surface pair, the data will be analyzed and curated in the MongoDB database right away.

\section{An Example of Substrate Screening via Hetero2d}
    \subsection{Materials Selection}
\begin{wrapfigure}[19]{r}{0.5\textwidth}
            \centering
            \vspace{-1.25\intextsep} 
            \includegraphics[width=0.5\textwidth]{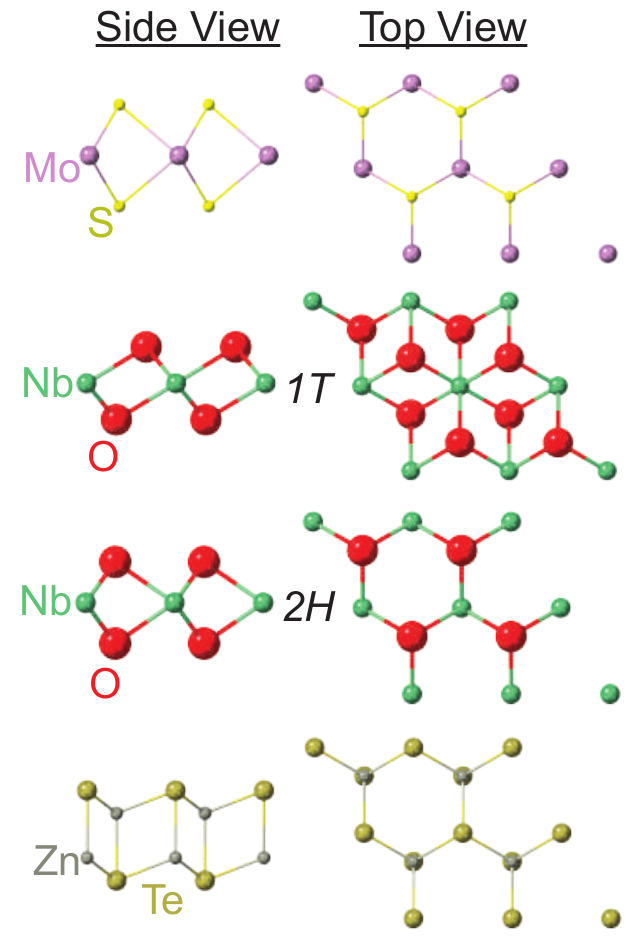}
            \vspace{-2\columnsep} 
            \caption{Structure models illustrating the 2D films crystal structure. Top view demonstrates the hexagonal symmetry of each 2D material. The $1T$ and $2H$ phase for NbO$_2$ are labeled to clarify the two phases.}
            \label{fig:2ds}
        \end{wrapfigure}

        To demonstrate the functionalities of the $Hetero2d$ package, we screened for suitable substrates for four 2D materials, namely $2H$-MoS$_2$, $1T$-NbO$_2$, $2H$-NbO$_2$~\cite{c2db}, and hexagonal-ZnTe~\cite{Torrisi2020}. The four 2D materials in consideration possess hexagonal symmetry as illustrated in Figure \ref{fig:2ds}. 
        
        MoS$_2$ was selected because there is a large amount of experimental and computational~\cite{Chen2013, Zhuang2013b, Yun2012, singh2015al2o3} data available in literature which we can use to validate the computed properties from our $Hetero2d$ workflow. The hexagonal-ZnTe~\cite{Torrisi2020}, $1T$-NbO$_2$, and $2H$-NbO$_2$~\cite{c2db} are yet to be synthesized. In addition, these particular 2D materials have diverse predicted properties see Table \ref{tab:2dProp}. It is noteworthy that hexagonal-ZnTe has been predicted to be an excellent CO$_2$ reduction photocatalyst~\cite{Torrisi2020}.  
    
        \begin{table}[!htbp]
            \centering
            \caption{The electronic properties and band gap of the four selected 2D materials used in this work. FM represents ferromagnetic.}
            \begin{adjustbox}{width=\textwidth}
                \begin{tabular}{|c|c|c|c|c|}
                \hline
                     2D Mat.        &  MoS$_2$        &  $1T$-NbO$_2$ & $2H$-NbO$_2$ & ZnTe \\
                \hline
                     Classification & Semiconductor   & FM~\cite{c2db}  & FM~\cite{c2db} & Semiconductor\\
                \hline
                     Band Gap (eV)  & 1.88~\cite{Gusakova2017} & 0.0~\cite{c2db} & 0.0~\cite{c2db} & 2.88~\cite{Torrisi2020} \\
                \hline
                \end{tabular}
            \end{adjustbox}
            \label{tab:2dProp}
        \end{table}
        \begin{table}[!htbp]
            \centering
            \caption{A list of matching substrate surfaces for the 4 2D materials given our heterostructure search criteria discussed in the next section.}
            \begin{adjustbox}{width=\textwidth}
                \begin{tabular}{|l|c|c|l|}
                \hline
                     2D Mat.     &  (111) Substrate &  (110) Substrate \\
                \hline
                     MoS$_2$     & Hf, Ir, Pd, Zr, Re, Rh & Ta, Rh, Sc, Pb, W, Y \\
                \hline
                     $1T$-NbO$_2$  & Ni, Mn, V, Nd, Pd, Ir, Hf, Zr, Cu & Rh, Ta, Sc, W \\
                \hline
                     $2H$-NbO$_2$  & Ni, Mn, Nd, Ir, Hf, Al, Te, Ag, Ti, Cu, Au & Ta, Sc, W, Y, Rh \\
                \hline
                     ZnTe        & Sr, Ni, Mn, V, Al, Ti, Cu & W\\
                \hline
                \end{tabular}
            \end{adjustbox}
            \label{tab:iface}
        \end{table}
       
        The properties of a 2D material can differ when placed on different miller-index planes for the same substrate. Thus, we investigated all unique low-index substrate surfaces (with $h$, $k$, $l$ equal to 1 or 0) for these 2D materials. A material available in the Materials Project (MP)~\cite{Ong2013}  database was considered a potential substrate if it satisfied all of the following criteria - a) is metallic, b) is a cubic phase, c) is single-element composition, d) has a valid ICSD ID~\cite{ICSD} (thus been experimentally synthesized), and e) has an $E_{above\ hull}<0.1$ eV/atom. There are 50 total substrates that satisfy the criteria above when queried from the MP database. 
        
        The bulk counterpart of each 2D material is also obtained from the MP database. We query the database for bulk materials that have the same composition as the 2D material and select the structure with the lowest $E_{above\ hull}$. SI Table 1-3 have additional reference information regarding all the optimized substrate slabs, 2D materials, and their bulk counterparts. SI Table 1 contains information about the Materials Project material\_id, $E_{above\ hull}$, ICSD ID, crystal system, and miller plane for the substrate surface. SI Table 2 contains information about the reference database ID, $\Delta E^{f}_{vac}$ (eV/atom), and crystal system for each 2D material and SI Table 3 contains information about the reference database id, $E_{above\ hull}$, E$_{gap}$, and the crystal system for the bulk counterpart of the 2D material. 
    
    \subsection{Symmetry-Matched, Lattice-Matched 2D-Substrate Heterostructures}
        In this study, we focus our search for 2D-substrate heterostructures to substrate planes with indices, $h$, $k$, $l$ as 0 or 1. The following studies focus on the heterostructures with the (111) and (110) substrate surfaces because we find that only these two miller planes have an appreciable number of heterostructures. The (001) substrate plane resulted in only one heterostructure. 
        
\begin{figure}[!htbp]
           \centering
           \includegraphics[width=0.5\textwidth]{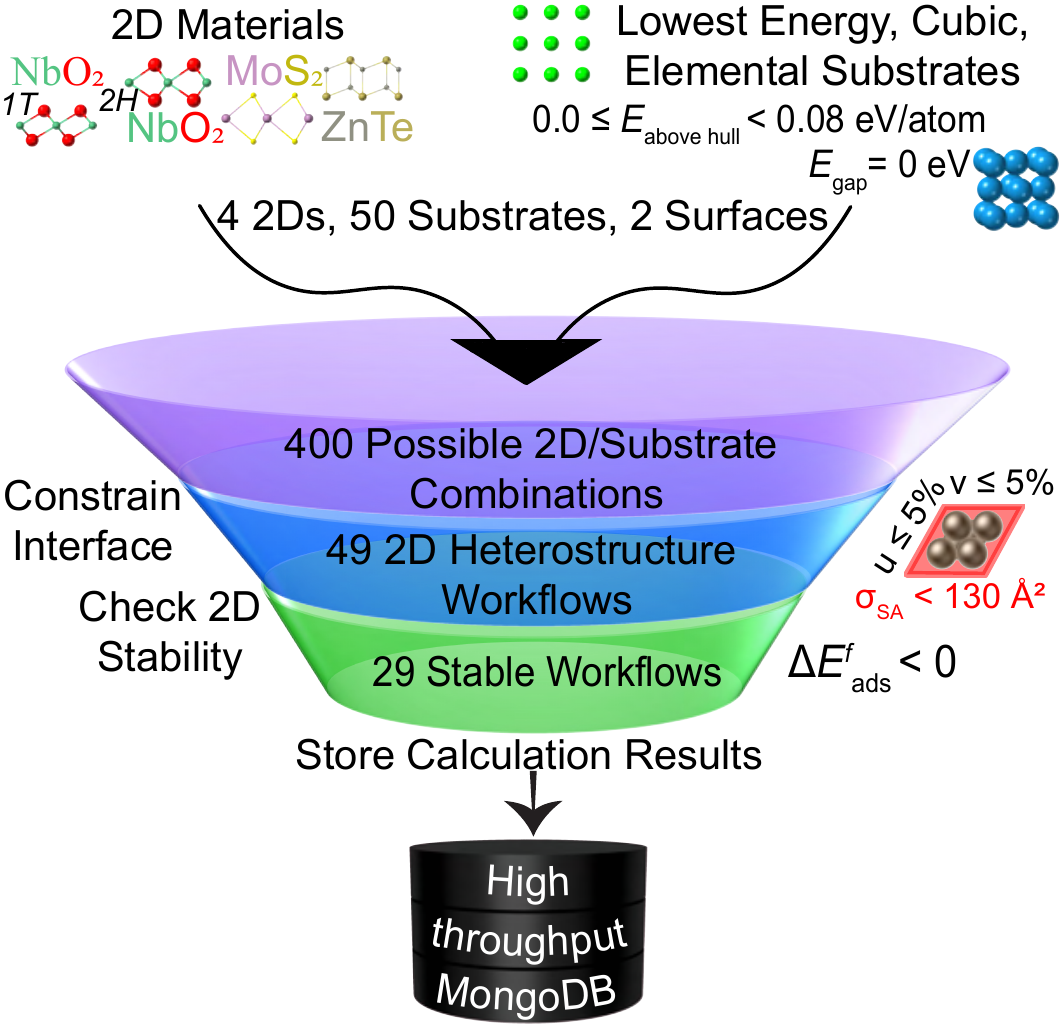}
           \caption{Schematic representing the materials selection process identifying stable 2D-substrate heterostructures using the $Hetero2d$ workflow. Tier 1 represents choosing 2D materials, substrates, and their surfaces. Tier 2 applies constraints on the surface area and lattice strain. Tier 3 shows the energetic stability of the heterostructures stored in the database.}
          \vspace{-0.25\intextsep} 
          \label{fig:Workflow}
        \end{figure}

        Restricting our search for 2D-substrate matches to only the (111) and (110) yields a total of 4 (\# of 2D materials) X 2 (\# of planes) X 50 (\# of substrates) = 400 potential 2D-substrate heterostructure combinations. As illustrated in Figure \ref{fig:Workflow}, after introducing our constraints for the surface area to be $< 130$ \AA\ and applied strain on the 2D material to be $ < 5$ \AA, a total of 49 2D-substrate heterostructure workflows are found. Table \ref{tab:iface} lists all metallic substrates matching each of the 2D materials given our heterostructure criteria.
        
        Of the total 49 workflows, 33 workflows correspond to the (111) substrate surfaces, and 16 workflows correspond to the (110) substrate surfaces. Generally, the (111) surface has more substrate matches than (110) surface due to the intrinsic hexagonal symmetry of the (111) surface that matches the hexagonal symmetry of the selected 2D materials. Each workflow generates between 2--4 2D-substrate heterostructure configurations for a given 2D-substrate surface pair, resulting in a total of 123 2D-substrate heterostructure configurations. Of those 2D-substrate heterostructures, 78 configurations, a total of 29 workflows stabilize the meta-stable 2D materials when placed upon the substrate slab. Additional details regarding these simulations can be found in section 4 of the SI.
        
    \subsection{Stability of Free-Standing 2D Films and Adsorbed 2D-Substrate Heterostructures}
        \begin{wrapfigure}[13]{r}{0.56\textwidth}
            \vspace{-1\intextsep} 
            \hspace*{-0.75\columnsep}\includegraphics[width=0.56\textwidth]{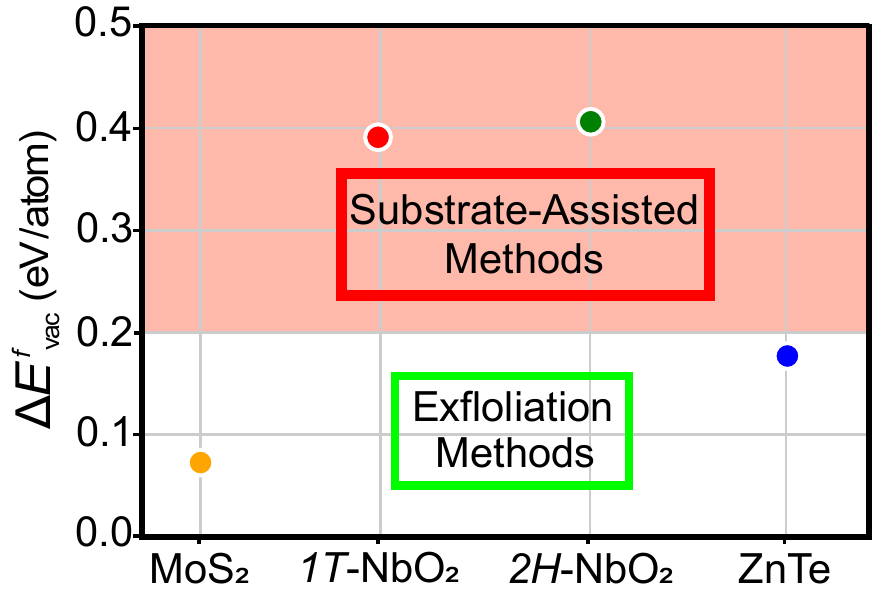}
            \vspace{-0.1\intextsep} 
            \caption{The \form\ for 2D- MoS$_2$ (\tikzcircle[gray, fill=orange]{2pt}), 
            $1T$-NbO$_2$ (\tikzcircle[gray, fill=red]{2pt}), 
            $2H$-NbO$_2$ (\tikzcircle[gray, fill=green]{2pt}), and ZnTe (\tikzcircle[gray, fill=blue]{2pt}). The \form\ is used to assess the thermodynamic stability of the free-standing 2D film with respect to its bulk counterpart. MoS$_2$ and ZnTe have relatively low \form\ while the $1T$ and $2H$ phase of NbO$_2$ have high \form.}
            \vspace{10\intextsep}
            \label{fig:form}
        \end{wrapfigure}
        Figure \ref{fig:form} shows the \form\ of the isolated unstrained 2D material with respect to their bulk counterpart. We find the \form\ for both MoS$_2$ and ZnTe are low, less than 0.2 eV/atom. Both the $1T$ and $2H$ phase for NbO$_2$ possess high \form, as shown by the red shaded region in Figure \ref{fig:form}, making substrate-assisted synthesis methods the most feasible method to synthesize these 2D films. The \form's in Figure \ref{fig:form} are consistent with prior computational~\cite{c2db, Torrisi2020} and experimental work~\cite{Lee2013}.
        
\begin{figure}[t!]
    	    \centering
            \vspace{-1\intextsep} 
    	    \includegraphics[width=\textwidth]{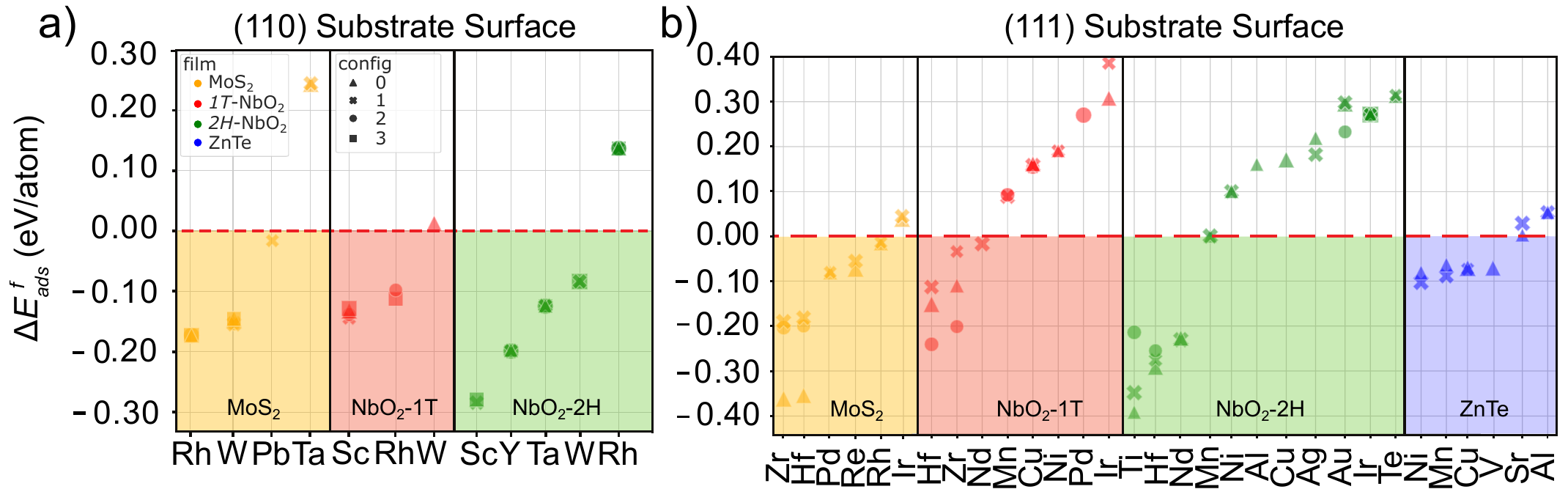}
            \vspace{-2\intextsep} 
    	    \caption{Adsorption formation energy, \ads, for the symmetry-matched, low lattice-mismatched (a) (110) and (b) (111) substrate surfaces. The rectangular symmetry of the (110) surface results in fewer matches while the hexagonal symmetry of the (111) substrate surface results in numerous matches within the given constraints on the surface area and lattice strain. Negative \ads\ values indicate stabilization of the 2D material. Each set of symbols (up to 4 points per substrate) represents the unique 2D-substrate configurations. }
            \vspace{-1\intextsep} 
    	    \label{fig:Eads}
    	\end{figure}

        Figures \ref{fig:Eads}a and \ref{fig:Eads}b show the \ads\ for the four 2D materials on the (110) and (111) substrate surfaces, respectively. The black lines in Figure 2 separate the 2D materials, while the shaded regions indicate stabilization of the 2D material on the substrate surface. When generating 2D-substrate heterostructure, the first challenge is finding a matching lattice between the 2D material and substrate surface. The next challenge is identifying "ideal" or likely locations to place the 2D material on the substrate surface to generate stable low-energy heterostructures. To reduce the large number of in-plane shifts possible for a given 2D-substrate heterostructure, we selectively placed the 2D material on the substrate slab by enumerating combinations of high-symmetry points (Wyckoff sites) between the 2D material and substrate slab stacking the 2D material on top of these sites $z$ \AA\ away from the substrate surface. Each unique 2D-substrate heterostructure configuration is represented by 0=$\triangle$, 1=\textbf{x}, 2=$\circ$, and 3=$\square$ in Figure \ref{fig:Eads}.
        
        The \ads\ on the (110) surface is shown in Figure \ref{fig:Eads}a. In the figure, 9 substrates stabilize the \ads\ of the 2D materials. The \ads\ appears to be correlated with the substrate where the 2D material is placed, however, there are not enough data points in Figure \ref{fig:Eads}a to distinguish the origin of this trend. Interestingly, when MoS$_2$ is placed on the (110) Ta substrate surface, the 2D material buckles which likely increases the \ads\ significantly above the other substrates. SI Figure 6 shows both configurations for MoS$_2$ on the (110) Ta substrate surface. There are an additional 5 2D-(110) substrate pairs that were studied but are not shown in Figure \ref{fig:Eads}a because the 2D materials/substrate interface becomes highly distorted/completely disintegrated. These cases are shown in SI Figure 4a and discussed in section 5 of the SI. 
    
        The (111) substrate surface matches for each 2D material are shown in Figure \ref{fig:Eads}b, where 15 substrates result in an \ads\ $<$ 0.  An additional 8 2D-substrate pairs, shown in SI Figure 4b, have 2D materials/substrate surfaces that are disintegrated and are discussed in section 5 of the SI.
        
        A correlation between the substrate surface and the \ads\ is more apparent for the (111) surface in Figure \ref{fig:Eads}b due to the increased number of 2D-substrate pairs. For MoS$_2$ on Zr and Hf, the triangle configurations have \ads\ significantly lower than the other configurations, see SI Figure 6 for structures of the three configurations. The lower \ads\ is correlated with smaller bond distances between the substrate surface and the 2D material. When the \ads\ is lower for these structures, we find that the $2h$ Wyckoff site of the 2D material is stacked on top of the $2a$ Wyckoff site of the substrate surface. The location of a 2D material on a substrate surface has previously been shown to influence the type of bonding present between the 2D material and substrate surface~\cite{Singh2014a,Zhuang2017}.
    
        The $1T$ phase of NbO$_2$ on Hf, Zr, and Ir substrates have an \ads\ difference between each configuration that is larger than other 2D-substrate pairs. The differences in \ads\ for $1T$-NbO$_2$ on Ir is partly due to some structural disorder of the 2D materials from the O atoms bonding strongly with the substrate surface, shown in SI Figure 7. For both Hf and Zr, the differences in \ads\ do not arise from structural disorder. The \ads\ of $1T$-NbO$_2$ on Hf and Zr are more strongly affected by the location of the 2D material on the substrate surface.
    
        $2H$-NbO$_2$ has two substrate surfaces, Ti and Au, where the \ads\ varies strongly with the configuration of 2D material on the substrate, unlike other 2D-substrate pairs for $2H$-NbO$_2$. $2H$-NbO$_2$ on Ti and Au have no structural distortions that explain the difference in \ads. For $2H$-NbO$_2$ on Ti, each configuration possesses different \ads\ arising from the unique placement of the 2D material on the substrate surface for each configuration. The strong bonding between the 2D material and substrate surface may be due to the affinity for Ti to form a metal oxide. SI Figure 8 shows each configuration for $2H$-NbO$_2$ on (111) Ti substrate surface. For $2H$-NbO$_2$ on Au, the circle configuration has a lower \ads\ due to the bottom layer of the $2H$-NbO$_2$ stacked directly on the top layer of the Au substrate surface. 
        
        The properties of MoS$_2$ have been studied both computationally and experimentally, where previous computational works~\cite{Zhuang2013b, Singh2015} have found similar values for the \form\ of MoS$_2$. Chen et. al. found that Ir bonds more strongly with the substrate surface than Pd~\cite{Chen2013}. This may explain the small structural modulations observed in our study for MoS$_2$ on the Ir (111) substrate surface but no such modulation is observed for MoS$_2$ on the Pd (111) substrate surface. Additionally, the $z$-separation distance between the 2D material and substrate surface found in this work agrees well with Chen et. al.'s values despite using a different functional. Our $z$-separation distances are within 0.05 \AA\ for Ir and 0.16 \AA\ for Pd~\cite{Chen2013}. 

    \subsection{Separation Distance of Adsorbed 2D Films on Substrate Slab Surfaces}
        The change in the thickness of the adsorbed 2D material may provide insight into the nature of bonding between the 2D-substrate heterostructures. For instance, vdW bonds are weak and thus typically result in minimal structural and electronic changes in the 2D material. Using our database, we determine the change in the thickness of post-adsorbed 2D materials from that of the free-standing 2D material. The thickness of the free-standing/adsorbed 2D material is computed first by finding the average $z$ coordinate of the top and bottom layer of the 2D material given by $\bar{d}_z = \sum\limits_{i=1}^n d^{top}_{i,z}/n - \sum\limits_{i=1}^m d^{bottom}_{i,z}/m$ where $d_{i,z}$ is the $z$ coordinate of the $i^{th}$ atom summed up to $n$ and $m$, the total number of atoms in the top and bottom layers, respectively. The thickness, obtained by taking the difference between the average thickness of the adsorbed 2D material from that of the free-standing 2D material, $\delta d$=$\bar{d}^{adsorbed}_z-\bar{d}^{free}_z$, with positive (negative) values corresponding to an increase (decrease) in the thickness of the adsorbed 2D material. 
        
\begin{figure}[!h]
            \centering
            \includegraphics[width=0.50\textwidth]{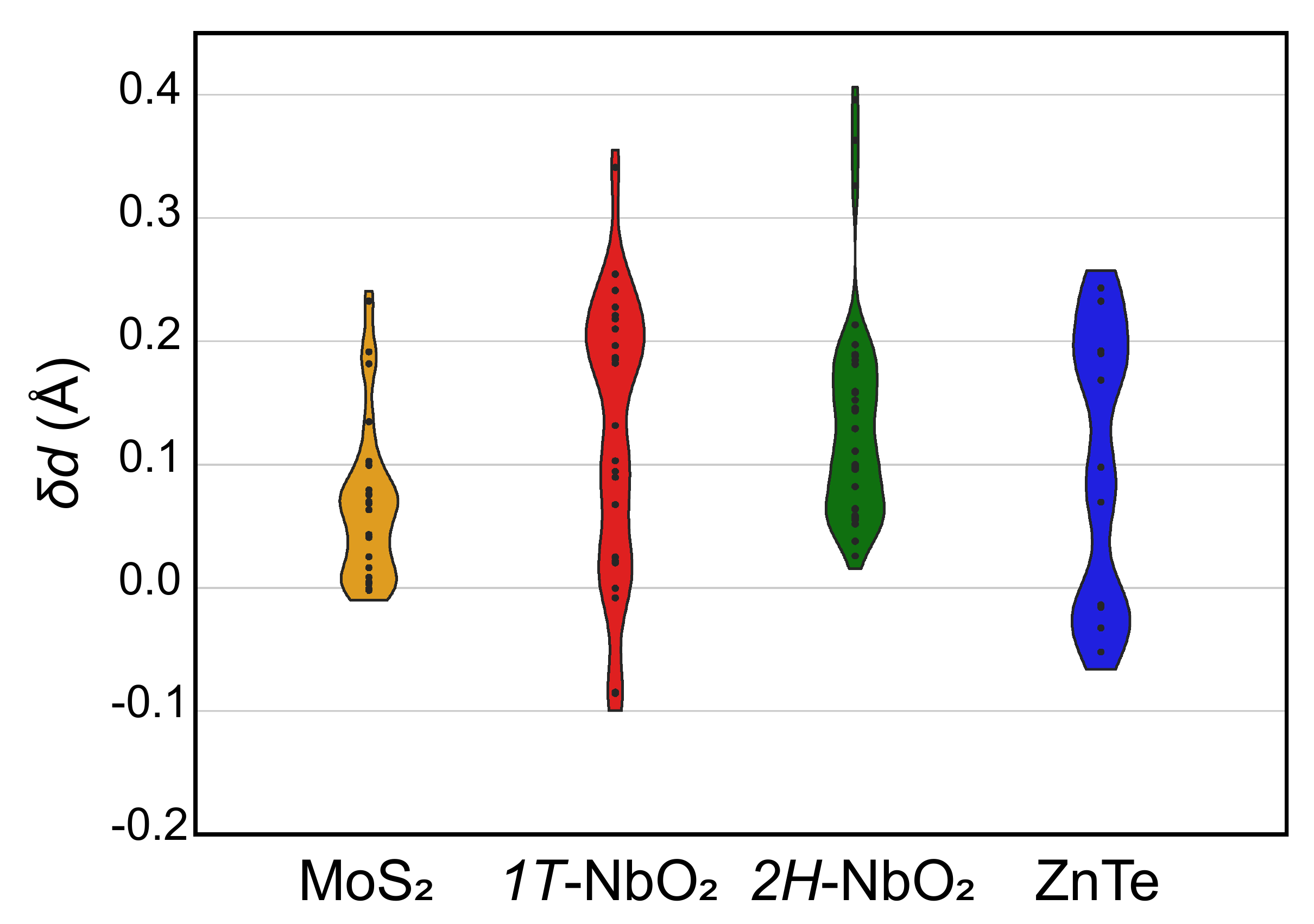}
            \caption{Each 2D material is separated spatially along the $x$-axis using a violin plot. The change in the 2D material's thickness, $\delta d$, for all substrates is plotted along the $y$-axis. A positive $y$-value indicates the 2D material's thickness has increased during adsorption onto the substrate slab. The width of the violin plot is non-quantitative from scaling the density curve by the number of counts per violin, however, within one violin plot, the relative $x$-width does represent the frequency that a 2D material's thickness changes by $y$ amount relative to the total number of data points in the plot.}
        \label{fig:Zdiff}
        \end{figure}

        Figure \ref{fig:Zdiff} illustrates the change in the thickness of the free-standing 2D material from that of the adsorbed 2D material for each 2D-substrate heterostructure. Typically for vdW type bonding, each atom should have minimum deviations from the free-standing 2D film due to the weak interaction between the adsorbed 2D material and substrate surface that characterizes vdW bonding. Figure \ref{fig:Zdiff} shows many of the 2D-substrate pairs have a significant change in the thickness of the 2D material that may indicate more covalent/ionic type bonding. The change in the thickness of the 2D material for the majority of the MoS$_2$-substrate configurations is minimal ($\textless$0.1 \AA) that may indicate weak interactions between the 2D material and substrate surface. Figure \ref{fig:Zdiff} indicates that for the majority of the adsorbed 2D materials, the substrates tend to induce an increase in the thickness of the adsorbed 2D material.

    \subsection{Charge Layer Doping of Adsorbed 2D Films}
        The $Hetero2d$ workflow package has a similar infrastructure as \textit{atomate} that allows our package to integrate seamlessly with the workflows developed within \textit{atomate}. These workflows enable us to expand our database by performing additional calculations such as Bader~\cite{Tang2009,Henkelman2006} charge analysis and high-quality density of states (DOS) calculations to assess charge transfer that occurs between the adsorbed 2D material and the substrate surface, changes in the DOS from the adsorbed and pristine 2D material, and changes in the charged state of the 2D-substrate pairs. 
   
        \begin{table}[h!]
            \centering
            \caption{Q$_x$ is obtained with Bader analysis and represents the average number of electrons transferred to/from (positive/negative) specific atomic layers with the initial number of electrons taken from the POTCAR. The first four columns are the electrons transferred to/from -- the Hf substrate atoms, Q$_{sub}$, the bottom layer of S atoms, Q$_{S_b}$, the Mo atoms, Q$_{Mo}$, and the top layer of S atoms, Q$_{S_t}$ for the adsorbed 2D-substrate heterostructure. The last three columns denote the charge transfer in the pristine MoS$_2$ structure. MoS$_2$ has an increased charge accumulation on the bottom layer of the 2D material from the substrate slab.}
            \begin{adjustbox}{width=3in}
               \begin{tabular}{|c|c|c|c|c|c|c|c|}
                 \hline
                   electrons & Q$_{sub}$ & Q$_{S_b}$ & Q$_{Mo}$ & Q$_{S_t}$ & Q$^{prist}_{S_b}$ & Q$^{prist}_{Mo}$ & Q$^{prist}_{S_t}$ \\
                 \hline
                   Q$_x$    & -0.11      & 1.10      & -1.03     &  0.57     & 0.60              & -1.20            & 0.60 \\
                \hline
               \end{tabular}
            \end{adjustbox}
            \label{tab:bader}
        \end{table}
        \begin{figure}[hb!]
           \centering
           \includegraphics[width=\textwidth]{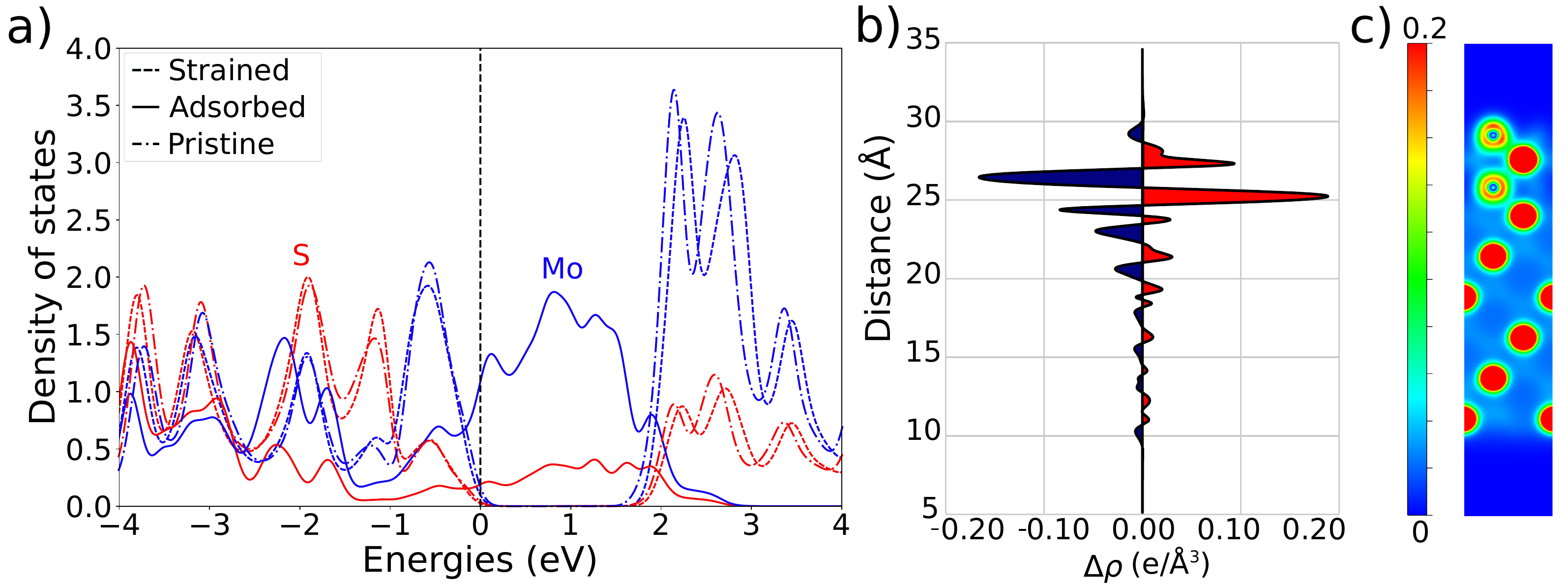}
           \caption{(a) The element projected density of states (DOS) where red and blue lines correspond to S and Mo states, respectively, for the isolated strained 2D material (dashed lines), the adsorbed 2D material (solid lines), and the pristine MoS$_2$ material (dashed-dotted lines). The Hf (111) substrate influences the DOS for MoS$_2$ causing a semiconductor to metal transition. (b) The $z$ plane-averaged electron density difference ($\Delta\rho$) for MoS$_2$ on Hf. Electron density difference is computed by summing the charge density for the isolated MoS$_2$ and isolated Hf then subtracting that from the charge density of the interacting MoS$_2$ on Hf system. The charge densities were computing with fixed geometries. The red and blue colors indicate electron accumulation and depletion in the combined MoS$_2$ on Hf system, respectively, compared to the isolated MoS$_2$ and isolated Hf atoms. (c) The charge density distribution for MoS$_2$ on (111) Hf substrate. The cross section is taken along the (110) plane passing through Mo, S, and Hf atoms. The charge density is in units of electrons/\AA$^3$.}
           \label{fig:DosChg}
        \end{figure}
        Most 2D materials are desirable due to their unique electronic properties. We selected MoS$_2$ on Hf (111) surface to demonstrate the capability of \textit{Hetero2d} in providing detailed electronic and structural information. Our Bader analysis illustrated in Table \ref{tab:bader} shows that there is charge transfer from the substrate to the bottom layer of the 2D material which is consistent with the findings presented by Zhuang et. al.~\cite{Zhuang2017} In Figure \ref{fig:DosChg}a, the DOS for the isolated un-strained, isolated strained, and adsorbed MoS$_2$ is shown where the black dashed line represents the Fermi level. There is a small shift in the DOS when comparing the un-strained and strained DOS for MoS$_2$. Comparing the DOS for the adsorbed MoS$_2$ to the other DOS for MoS$_2$, there is a significant change in the DOS. We can see that the substrate influences the DOS of MoS$_2$ when placed on the Hf (111) surface causing a semiconductor to metal transition of the MoS$_2$. This change in the DOS is consistent with the Bader analysis that indicates electron doping of the MoS$_2$ material occurs which would result in changes in the DOS. Figure \ref{fig:DosChg}b shows the redistribution of charge due to the interaction of the 2D material and substrate surface where red and blue regions indicate charge accumulation (gaining electrons) and depletion (losing electrons) of the combined system due to the interaction between MoS$_2$ and Hf. The charge density difference is computed as the difference between the sum of the isolated MoS$_2$ and isolated Hf substrate slab from that of the combined MoS$_2$ on Hf system . Figure \ref{fig:DosChg}c is the charge density of the combined MoS$_2$ on Hf system along the (110) plane. Thus, the electronic properties of MoS$_2$ are dramatically affected by the substrate. \textit{Hetero2d} can analyze the substrate induced changes in the electronic structure of 2D materials. This will lead to a fundamental understanding and engineering of complex interfaces.

\section{Conclusions} 
    In summary, we have developed an open-source workflow package, $Hetero2d$, that automates the generation of 2D-substrate heterostructures, the creation of DFT input files, the submission and monitoring of computational jobs on supercomputing facilities, and the storage of relevant parameters alongside the post-processed results in a MongoDB database. Using the example of four candidate 2D materials and low-index planes of 50 potential substrates we demonstrate that our open-source package can address the immense number of 2D material-substrate surface pairs to guide the experimental realization of novel 2D materials. Among the 123 configurations studied, we find that only 78 configurations (29 workflows) result in stable 2D-substrate heterostructures. We exemplify the use of $Hetero2d$ in examining the changes in thickness of the adsorbed 2D materials, the Bader charges, and the electronic density of states of the heterostructures to study the fundamental changes in the properties of the 2D material post adsorption on the substrate. $Hetero2d$ is freely available on our GitHub website under the GNU license along with example jupyter notebooks. 
  
\section{Acknowledgements}
    The authors thank start-up funds from Arizona State University and the National Science Foundation grant number DMR-1906030. This work used the Extreme Science and Engineering Discovery Environment (XSEDE), supported by National Science Foundation grant number TG-DMR150006. The authors acknowledge Research Computing at Arizona State University for providing HPC resources that have contributed to the research results reported within this paper. This research also used resources of the National Energy Research Scientific Computing Center, a DOE Office of Science User Facility supported by the Office of Science of the U.S. Department of Energy under Contract No. DE-AC02-05CH11231. The authors acknowledge Akash Patel for his dedicated work maintaining our database and API. We thank Peter A. Crozier for their valuable discussions and suggestions. 

\section{Supporting Information}
    Supporting information provides additional descriptions, figures, and tables supporting the results described in the main text.

\section{Data Availability}
    The results reported in this article and the workflow package can be found on our github website \href{https://github.com/cmdlab/Hetero2d}{Hetero2d}. 
\bibliography{References}

\end{document}